\documentstyle[PASJadd,epsf,psfig]{PASJ95}

\draft

\newcommand{\vol}{}
\newcommand{\no}{}
\newcommand{\lett}{}
\newcommand{\spage}{}
\newcommand{\rdate}{}
\newcommand{\adate}{}

\begin{document}

\title{ASCA Observations of Two Ultra-Luminous Compact X-Ray Sources
in the Edge-on Spiral Galaxy NGC~4565}

\author{
Tsunefumi {\sc Mizuno}, Tomohisa {\sc Ohnishi}, Aya {\sc Kubota},
Kazuo {\sc Makishima}, \\
and Makoto {\sc Tashiro} \\
{\small \it Department of Physics,  University of Tokyo,
7-3-1 Hongo, Bunkyo-ku, Tokyo 113-0033}\\
{\small \it E-mail(TM): mizuno@amalthea.phys.s.u-tokyo.ac.jp}\\
}

\abst{
The edge-on spiral galaxy NGC~4565 was observed for $\sim$ 35 ks
with ASCA in the 0.5--10~keV energy band.
The X-ray emission was dominated by two bright sources,
which can be identified with two point-like X-ray sources seen in the
ROSAT HRI image.
The observed 0.5--10 keV fluxes of these sources,
$\rm 1.7\times 10^{-12}~erg~s^{-1}~cm^{-2}$ and
$\rm 0.7\times 10^{-12}~erg~s^{-1}~cm^{-2}$,
imply bolometric luminosities of
$\rm 1.0\times 10^{40}~erg~s^{-1}$ and $\rm 4 \times 10^{39}~erg~s^{-1}$,
respectively.
They exhibit similar spectra, which can be explained by emission
from optically thick accretion disks
with the inner disk temperature of 1.4--1.6 keV.
One of them, coincident in position with the nucleus, shows too low absorption
to be the active nucleus seen through the galaxy disk.
Their spectra and high luminosities suggest that they are
both mass accreting black hole binaries.
However the black-hole mass required by the Eddington limit is rather high
($\geq$ 50 $M_{\odot}$), and the observed disk temperature is too high
to be compatible with the high black-hole mass.
Several attempts are made to solve these problesms.
}
\kword{
Black hole physics --- Galaxies: individual (NGC~4565) ---
Galaxies: spiral --- Galaxies: X-rays}

\maketitle
\thispagestyle{headings}

\section{Introduction}

It has long been known (e.g., Fabbiano 1988)
that a fair number of nearby spiral galaxies host
extremely luminous X-ray sources with apparently point-like appearances,
or ``ultra-luminous compact X-ray sources'' (hereafter ULXs).
Their luminosities can reach 
$\rm \sim 10^{40}~erg~s^{-1}$ (e.g., Read et al.\ 1997),
exceeding the Eddington limit for a 1.4~$M_{\odot}$
neutron star almost by two orders of magnitude.
A general indication is
that they are indeed mass accreting compact objects
associated with the host galaxies,
rather than background or foreground contaminants (e.g., Fabbiano 1989).
The extremely high luminosities of these objects are
thought to indicate some extraordinary conditions of them.

Clarifying the nature of ULXs is of great importance,
because they strongly influence our understanding
of the X-ray emission from normal spiral galaxies.
For example, a ULXs located close to the galaxy center would be
mistaken for an active galactic nucleus (AGN) of low luminosity.
However, the absence of ULXs in the Milky Way and M31 has hampered
clear identification of their nature.

Through observations with ASCA (Tanaka et al.\ 1994),
0.5--10~keV X-ray spectra have been accumulated on a fair number of ULXs
(Petre et al.\ 1994; Takano et al.\ 1994; Reynolds et al.\ 1997;
Okada et al.\ 1998; Uno 1997).
Although these ASCA spectra exhibit a fair degree of variety,
some of them have been fitted successfully with so called multi-color
disk blackbody (MCD) model.
Such examples include the center source (X-8) of M33 (Takano et al.\ 1994),
the brightest source (source 1) in IC~342 (Okada et al.\ 1998),
and the source X-6 in M81 (Uno 1997).
The MCD model describes optically-thick multi-temperature emission from
a standard  accretion disk (Shakura, Sunyeav 1973) around a black hole.
It can be characterized by the highest disk temperature, $T_{\rm in}$,
and the size of the accretion disk
(Mitsuda et al.\ 1984; Makishima et al.\ 1986).

Although these ASCA results provide an impotant clue to the
understanding of ULXs in terms of mass-accreting black holes,
in some cases there remains a fundamental problem (Okada et al.\ 1998)
that the measured values of $T_{\rm in}$ are uncomfortably high (1.0--1.8~keV).
In order to further examine this issue,
it is essential to enlarge the sample of such spectral measurements.
Accordingly, we here analyze the ASCA data
of the nearby edge-on spiral galaxy NGC~4565.
We show that it contains two ULXs, one nearly coincident in projection
with its nucleus.
Furthermore, we confirm that the spectra of these two sources can be
described well by the MCD model with rather high disk temperature.

\section{Observations and Results}

\subsection{Observation and Data Reduction}

NGC~4565 is a spiral galaxy with an almost perfectly edge-on geometry
of inclination $\sim$86$^{\circ}$ (Hummel et al.\ 1984 ).
It is located at a high galactic latitude ($b = 86.4^{\circ}$)
and an estimated distance of $\rm \sim 9.7~Mpc$ by the
Tully-Fisher relation (Tully 1988).
Other distance indicators,
such as the globlar cluster luminosity function, 
planetly nebula luminosity function, 
and surface brightness fluctuations, 
all indicate similar distance of $\sim$ 10 Mpc
(Fleming et al.\ 1995; Simard \& Pritchet.\ 1994;
Jacoby et al.\ 1996).

NGC~4565 was observed by ROSAT three times, once with the HRI and
twice with the PSPC (Volger et al.\ 1996).
The archival ROSAT HRI image shown
in figure 1a
is dominated by two point-like X-ray sources,
one about $0.\hspace{-2pt}'8$ above the galaxy disk
while the other coincident in position with the galaxy nucleus
within the ROSAT position accuracy of $\sim$4$''$
(Volger et al.\ 1996; Rupen\ 1991).

We observed NGC~4565 with ASCA on 1994 May 28.
The SIS (Solid State Imaging Spectrometer; Burke et al.\ 1994;
Yamashita et al.\ 1997) data were acquired in 2CCD/FAINT mode,
while the GIS (Gas Imaging Spectrometer; Ohashi et al.\ 1996;
Makishima et al.\ 1996) data were taken in the standard PH mode.

The SIS data were selected using the following criteria:
a) time after passage through the South Atlantic Anormaly
(SAA) was greater than 1~minute;
b) the object was at least $10^{\circ}$ above the night Earth's limb;
c) the object was at least $20^{\circ}$ above the bright Earth's limb;
d) the cutoff rigidity (COR) of cosmic rays  was greater than 6~GeV/c; and
e) time after day night transitoin was greater than 100~s.
We also removed hot and flickering pixels.
The GIS data were selected using the following criteria:
a) time after SAA was greater than 1~minute;
b) the object was at least $10^{\circ}$ above the Earth's limb; and
c) the cutoff rigidity (COR) was greater than 6~GeV/c.
We also used the standard rise-time rejection and spread discrimination
to remove particle events.
After applying these criteria, we obtainded $\rm \sim31.5~ks$ of good SIS data,
and $\rm \sim 35.5~ks$ of good GIS data.
The galaxy was clearly detected,
at counting rates of $\rm 0.08~c~s^{-1}$ (in 0.5--10~keV) per SIS detector,
and $\rm 0.05~c~s^{-1}$ (in 0.7--10~keV) per GIS detector.


\subsection{X-Ray Images}

Figure 1b shows
the ASCA SIS (SIS0 plus SIS1) image of NGC~4565 after
correcting the attitude data for known temperature effects.
The ASCA image comprises two emission peaks with a separation of
$\sim 0.\hspace{-2pt}'8$,
presumably corresponding to the two sources visible
in the HRI image (figure 1a).

To confirm the presence of the two sources in the ASCA image,
we have projected the SIS events inside the rectangle of
figure 1b
onto its longer side.
The derived one-dimensional X-ray profiles are shown
in figure 2,
in three representative energy bands.
Each profile can be fitted well by the projected Point Spread Function (PSF)
of the X-ray image convolved with the two point sources,
of which the locations are fixed to those deterimined with the ROSAT HRI.
We assumed constant background and determined the intensities of the two sources in the three
energy bands, as tabulated
 in table 1.
We can see that intensity ratios of the two sources observed by ASCA
are energy independent within the statistical errors.
The discripancy between the ASCA and ROSAT HRI ratios
may be due to time valiability.



\subsection{Summed Energy Spectra}

Now that the two sources have been confirmed to have similar spectra,
we tentatively analyze their spectra together as a first-cut analysis.
We hence accumulated the SIS and the GIS events over
circular regions of radii $4'$ and $6'$ respectively, both
centered on the off-center (brighter) source,
and obtained the spectra shown
in figure 3a.
We subtracted background spectra accumulated using a source free region
of the same dataset for
the SIS, and blank-sky data for the GIS.
We fitted these SIS/GIS spectra simultaneously with a common model;
either a power-law, a thermal Bremsstrahlung,
a plasma emission (Masai 1984), a blackbody, or an MCD model.
For the spectral fitting, we used the response function
for a point source located at the off-center source.

The fit results are summarized
in table 2.
Thus, the blackbody model was unsuccessful.
The power-law model gives a much better fit with $\chi^2/\nu = 1.44$,
but the probability of this fit being acceptable is less than 1\%.
Residuals to the power-law fit indicate
that the observed spectra are more convex than a single power-law;
even if the two sources have different spectral slopes,
their sum would not exhibit such a convex shape.
We therefore rule out the power-law fit.

The remaining three models all give acceptable fits to the data.
However, the plasma emission fit requires a very low metallicity;
therefore it is basically the same as the Bremsstrahlung fit.
The Bremsstrahlung model, in turn, requires an intrinsic absorption
much exceeding the galactic line-of-sight column density of
$N_{\rm H} = 1.3 \times 10^{20}~{\rm cm}^{-2}$.
Such a moderately absorbed Bremsstrahlung model is known to
empirically approximate optically-thick emission arising from
accreting non-magnetized compact objects (Makishima et al.\ 1989),
rather than having its own physical meaning.
We therefore regard the Bremsstrahlung fit
as essentially equivalent to the MCD fit,
which is also acceptable.
The MCD fit has yielded $T_{\rm in} \sim 1.4~{\rm keV}$,
and the 0.5--10~keV flux of the sum of the two sources becomes
$\rm 2.3\times 10^{-12}~erg~s^{-1}~cm^{-2}$.
The flux is thought to be dominated by the two sources,
because the contribution from the underlying ordinary X-ray binaries
are estimated to be $<10$\% by assuming that
the $f_{\rm X}$/$f_{\rm B}$ ratio is the same as that of M31
(Makishima et al.\ 1989),
where $f_{\rm X}$ represents the X-ray flux and $f_{\rm B}$ represents
the optical B-rand flux.

Although the single MCD fit is acceptable,
the observed spectra may contain an additional hard component.
For example, if the two sources are low-mass X-ray binaries
(LMXBs; close binaries involving non-magnetic neutron stars),
we expect to detect a blackbody hard component of temperature $\sim$2~keV,
arising from the neutron-star surface (Mitsuda et al.\ 1984).
To examine this possibility,
we refitted the SIS and GIS spectra with the MCD model plus a blackbody model,
with the blackbody temperature fixed at 2.0~keV to ensure a stable fitting.
The MCD parameters and the absorption were allowed to float again.
However the data did not require the blackbody component,
with its 0.5--10~keV flux being $<33$\% (90\% confidence)
of the total 0.5--10~keV flux.
In short, the obtained spectra are considerably softer than those of LMXBs.
The result of Bremsstrahlung fit given
in table 2
also supports this interpretation,
because the Bremsstrahlung approximation to the ASCA spectra of LMXBs
usually give a considerably higher temperature of $\rm 8 \sim 13~keV$.

We also repeated the MCD fitting by replacing the blackbody component
by a hard power-law with its photon index fixed at 2.2:
this simulates the spectra of black-hole binaries
in the high (or soft) state (Tanaka, Lewin 1995).
The power-law was not required either,
with its 0.5--10~keV flux being $<35$\% (90\% confidence)
of the observed flux in the same range.



\subsection{Spectra of Individual Sources }

As a further investigation,
we attempted to estimate the spectra of the two sources individually.
We therefore fitted the same SIS and GIS spectra
jointly with a sum of two MCD components having
separate temperatures, separate normalizations, and separate column densities,
which are all left free to vary.
We imposed additional constraints that each MCD
component should correctly reproduce the three-band coarse spectrum
of the corresponding source, wich was produced 
by converting the count rates in table 1.
The response function to represent the three-band spectra was made
by scaling and rebinning the response function used in section 2.2.
As a result, we fitted four spectra
(whole spectrum of SIS, that of GIS,
three-band SIS spectrum for the off-center sources,
and that for the center source)
and determine the model parameter
to minimize the total chi-square.
In other words, we performed a simultaneous fitting
to the spectral and imaging data sets.

This composite model have given an acceptable fit
with an overall chi-squared of 147.7 for 124 degree of freedom,
and the obtained results are summarized
in table 3.
Thus, the two sources exhibit the same disk temperature within errors,
which in turn agree with that derived
in table 2.
This justifies our analysis performed in section 2.3.
Neither source exhibit detectable absourption, again in agreement
with table 2.
We repeated the same fitting using different model combinations.
When the off-center source spectrum is represented by a power-law,
the overall fit becomes unacceptable ($\chi^{2}$/$\nu$=159.1/124).
In contrast, the center source spectrum was described equally well
($\chi^{2}$/$\nu$=143.9/124)
when its MCD model is replaced by a power-law model of photon index
$\Gamma$ = 1.55$^{+0.28}_{-0.22}$.
However, the absorption associated with this power-law model for the
center source remained rather low ($\le$2$\times$$10^{21}~{\rm cm}^{-2}$),
with the MCD parameters for the off-center source remained unchanged 
within the statistical errors.


\section{Discussion}

Using ROSAT and ASCA,
we have detected two point-like luminous X-ray sources
in the edge-on spiral galaxy NGC~4565.
Their spectra have been described successfully by the MCD model
with disk temperatures $T_{\rm in} = 1.4-1.6~{\rm keV}$,
although that of the center source can also be described with a power-law
of $\Gamma$ = 1.55.

\subsection{The Off-Center Source}

The off-center source is the brighter of the two.
It locates $\rm \sim 2~kpc$ above the galaxy disk.
Since this exceeds the typical scale height ($\rm \sim 0.2~kpc$)
of the X-ray source distribution in Our Galaxy,
the source may be suspected to be a background AGN or a foreground object.
However, a chance probability to find an X-ray source of
0.5--10~keV flux exceeding $\rm 1 \times 10^{-12}~erg~s^{-1}$
in a particular sky region encompassing NGC~4565, e.g.,
$2' \times 10'$ in size,
is only $\sim 0.3$\%, as calculated from the $\log N$--$\log S$ distribution
for extragalactic X-ray sources
(e.g., Ueda et al.\ 1998).
Furthermore, none of AGNs are known to exhibit an MCD-type spectrum
that has a mildly concave shape in a logarithmic plot.
The chance probability of this source being
a forground Galactic objtect is also quite low,
because of its location close to the Galactic north pole.
We therefore conclude that this source is associated with NGC~4565.

Applying bolometric correction via the MCD model
and employing the 9.7~Mpc distance,
the bolometric luminosity of the off-center source becomes
$\rm 2.0 \times 10^{40}~erg~s^{-1}$ if the emission is isotropic,
or $1.0 \times 10^{40}~(\cos i_{\rm o})^{-1}$~erg~s$^{-1}$
if assuming a flat disk geometry with an inclination $i_{\rm o}$.
These values much exceed the Eddington limit for a
$1.4~M_{\odot}$ neutron star.

The high luminosity and a relatively large distance above the galaxy disk
suggest that this source belongs to a globular cluster in NGC~4565.
However, we would then need some
$\sim 100$ or more LMXBs residing in one or a few globular clusters,
with each LMXB radiating at a near-Eddington luminosity.
Such a high concentration of luminous LMXBs is
never observed in the Milky Way or M31.
Furthermore, due to the averaging effect,
we would then observe a typical LMXB spectrum,
with nearly equal luminosities in the
MCD and blackbody components (Mitsuda et al.\ 1984).
This contradicts our results derived in section 2.3, and hence unlikely.
We therefore conclude that the off-center souce is a ULXs.

\subsection{The Center Source}

At a distance of 9.7~Mpc,
the bolometric luminosity of the center source is
$\rm 8 \times 10^{39}~erg~s^{-1}$ for an isotropic emission,
or $4 \times 10^{39}~(\cos i_{\rm c})^{-1}$~erg~s$^{-1}$
for a disk-like emission, where $i_{\rm c}$ is the inclination of this object.
This also exceeds very much the Eddington limit for a neutron star.

Then, together with its positional coincidence
(within $\rm \sim 0.2~kpc$) with the nucleus of NGC~4565,
the simplest account of the souce would be a low-luminosity AGN of NGC~4565.
However, the clear detection of this srouce with the ROSAT HRI
(figure 1a)
and the ASCA spectral results
(table 3)
consistently indicate that the absorbing column density
to this source is quite low ($\le 0.5 \times 10^{21}$~cm$^{-2}$
in terms of the MCD model).
Even if we assume a power-law model for the center source,
$N_{\rm H}$ still remains rather low
($\le 2.0 \times 10^{21}$~cm$^{-2}$).
On the other hand, judging from the inclination of 86$^{\circ}$
for NGC~4565 (Hummel et al.\ 1984),
and scaling the absorption column to Our Galaxy center of
$\sim 5 \times 10^{22}$~cm$^{-2}$ (Predehl et al.\ 1994),
the column density along the galaxy disk to the nucleus of NGC~4565
would amount at least to $\rm \sim 1 \times 10^{22}~cm^{-2}$.
Such a large absorption is ruled out by the data,
therefore the center source is not likely to be a low-luminosity AGN.
The low absorption also rules out securely this source being a bakcground AGN.
Judging from its particular position,
this source is not likely to be a foreground object, either.
Furthermore, the same argument as conducted in the previous subsection
makes this source unlikely to be an assembly of luminous LMXBs.
We therefore suggest
that the center source is a second ULX in NGC~4565,
located at the near-side edge of the disk of the galaxy.

\subsection{The Nature of the Two Sources and Associated Problems}

Taking it for granted that the two sources are both ULXs,
the clue to their nature may be provided by their spectra.
These objects cannot be very luminous Crab-like supernova remnants,
since the power-law fit failed to describe
the spectrum of the off-center source (\S~2.4),
and the photon index of $\Gamma=1.55$ for the center source
is inconsistent with the typical Crab-type slope of $\Gamma \sim 2.0$.
As already discussed, the interpretation as an assembly of LMXBs
is incompatible with the measured spectra.
An X-ray pulsar with a significant radiation beaming is also unlikely,
because luminous Galactic and Magellanic X-ray pulsars exhibit
significantly flatter continua in the ASCA band,
with typical photon indices in the range 0.5--1.2 (Nagase 1989).

In contrast, we have reproduced their 0.5--10 keV spectra successfully
by the MCD model (or its empirical approximation
by an absorbed thermal Bremsstrahlung model).
Furthermore, the MCD model is applicable to the ASCA spectra
of three other ULXs as mentioned in section 1.
These results indicate that the X-ray emission from thse two ULXs,
and possibly from some other ULXs too,
originate from optically-thick accretion disks in these objects.
These ULXs are therefore inferred to be mass-accreting black holes in the
high state,
wherein the MCD emission from the accretion disk dominates in the X-ray band.
However, as discussed below,
this interpretation involves two serious problems.

An immediate problem associated with the black-hole interpretation of ULXs is
that the black-hole mass required to satisfy the Eddington limit is quite high.
In fact, for the off-center and the center sources
to be radiating below the Eddington limit,
their black-hole mass has to be
\begin{equation}
M_{\rm o}^{\rm Ed} > 73 \; (\cos i_{\rm o})^{-1}~M_\odot~,
~~~
M_{\rm c}^{\rm Ed} > 29 \; (\cos i_{\rm c})^{-1}~M_\odot~~,
\label{eq:M_edd}
\end{equation}
respectively.
Black holes as massive as  $M_{\rm o}^{\rm Ed}$ have never been observed,
and invoking such objects may contradict
current understanding of the stellar evolution.

The other, more delicate problem is
that the measured values of $T_{\rm in}$ of the two ULXs are rather high
compared to those of Galactic and Magellanic black-hole binaries
(typically 0.5--1.2 keV; Tanaka, Lewin 1995).
As first pointed out by Okada et al. (1998) with respect to the ULX in IC~342,
this implies a serious self-inconsitency in the black-hole interpretation of
ULXs;
below, we describe this issue.

The bolometric luminosity of an MCD emission may be decribed as
(Mitsuda et al. 1984; Makishima et al. 1986)
\begin{equation}
L_{\rm bol} = 4 \pi( R_{\rm in}/\xi)^2 \sigma (T_{\rm in}/\kappa)^4~~,
\label{eq:L_bol}
\end{equation}
where $R_{\rm in}$ is the innermost radius of the optically-thick accretion
disk,
$\sigma$ is the Stefan-Boltzmann constant,
$\kappa \sim 1.7$ (Shimura, Takahara 1995) is
ratio of the color temperature to the effective temperature,
and $\xi = (3/7)^{1/2}(6/7)^3 = 0.41$ (Kubota et al. 1998) is a correction
factor.
By substituting the bolometric luminosities of the two sources,
and assuming $\kappa=1.7$ and $\xi=0.41$,
we may solve equation (2) to obtain
$R_{\rm in} = 173^{+21}_{-16} \ (\cos i_{\rm o})^{-1/2}$ km
and $R_{\rm in} = 83^{+26}_{-22} \ (\cos i_{\rm c})^{-1/2}$ km
for the off-center and the center sources, respectively.

After previous works (e.g. Dotani et al. 1997),
we may identify $R_{\rm in}$ with the radius
inside which stable Kepler orbits no longer exist.
This radius becomes $3 \; R_{\rm S}$ for a non-smpinning black
hole of mass $M$,
where $R_{\rm S} = 2GM/c^2 = 9.0(M/M_\odot)$ km is the Schwarzschild radius,
with $G$ the constant of gravity and $c$ the light speed.
We then obtain $M_{\rm o}= 19.2^{+2.4}_{-1.8}~(\cos i_{\rm o})^{-1/2}~M_\odot$
for the off-center source,
and $M_{\rm c}= 9.2^{+2.9}_{-2.5}~(\cos i_{\rm c})^{-1/2}~M_\odot$
for the center source.
These values considerably fall short of the mass lower limits
imposed by equation (\ref{eq:M_edd}),
even if taking the most favourable case of $i_{\rm o}=i_{\rm c}=0$.
This severe self-inconsistency arises
because $T_{\rm in}$ is too high, and hence $R_{\rm in}$ is too small,
for the large black-hole mass required by the high luminosities.
Essentially the same problem has been reported
by Okada et al. (1998) on the brightest ULX in IC~342.

\subsection{Possible Solutions to the Problems}

We here attempt to solve the two problems raised in \S~3.3,
that the inferred black-hole mass is too high,
and that the disk is too hot.
One simple way around these problems is to assume
that the black-hole has a reasonable mass, e.g. 10--20 $M_\odot$,
and the observed high X-ray flux is due to radiation beaming toward us.
This hypotesis has often been employed by various authors
in a rather ad-hoc way,
when discussing the nature of ULXs.
However, none of them have successfully presented mechanisms
to produce such a radiation beaming.
Consequently, we do not appeal to this solution.

Another obvious solution to the issue is to presume
that the employed distance to NGC~4565, $D=9.7$ Mpc,
was significantly over-estimated.
However, we do not have a large degree of freedom of changing $D$,
since various distance estimates consistently yield $D \sim 10$ Mpc
as already mentioned in \S~2.1.
Here, let us take a rather extreme assumption of
$D=4.9$ Mpc instead of 9.7 Mpc,
and also assume $i_{\rm o}=i_{\rm c}=0$.
Then, the bolometric luminosities of the two sources become
$2.6 \times 10^{39}$ ergs s$^{-1}$ and $1.0 \times 10^{39}$ ergs s$^{-1}$,
and hence the mass limits of equation (\ref{eq:M_edd})
become
\begin{equation}
M_{\rm o}^{\rm Ed} > 18~M_\odot~, 
~~~M_{\rm c}^{\rm Ed} > 7~M_\odot~~.
\end{equation}
These values may be reasonable for stellar-mass black holes.
Thus, the issue of too high a black-hole mass might be solved by assuming
that the distance to NGC~4565 is over-estimated by a factor of 2,
and that the two objects are both face-on systems.

Under these assumptions and from equation (\ref{eq:L_bol}),
we obtain $R_{\rm in} = 86.3^{+10.7}_{-8.0}$~km
and $R_{\rm in} = 41.5^{+13.1}_{-11.2}$~km
for the off-center and the center sources, respectively,
because $R_{\rm in}$ is directly proportional to $D$.
By identifying these again with $3 \; R_{\rm S}$,
we then obtain $M_{\rm o}= 9.6^{+1.2}_{-0.9}~M_\odot$
for the off-center source,
and $M_{\rm c}= 4.6^{+1.5}_{-1.2}~M_\odot$ for the center source.
These values still remain inconsistent with the mass lower limits
imposed by equation (3).

To solve the remaining inconsistency, we notice
that the estimates of $\kappa$ or $\xi$ may be modified,
because these values must be subject to considerable uncertainties.
For this purpose, let us consider Cygnus X-1,
the most well studied black-hole binary
that is thought to have D=2.5 kpc and $i \sim 30^{\circ}$
(Dotani et al. 1997).
By employing $T_{\rm in}=0.43$ keV
and $L_{\rm bol}=2.4 \times 10^{37}$ ergs s$^{-1}$
measured with ASCA (Dotani et al. 1997),
together with $\kappa=1.7$ and $\xi=0.41$,
and equating again $R_{\rm in}$ with $3 \; R_{\rm S}$,
we obtain the black-hole mass of Cygnus X-1 to be $\sim 10~M_\odot$.
This agrees well with the optically estimated mass,
$10.1^{+4.6}_{-5.3}~M_\odot$ (Herreo et al. 1995).
From this result,
we infer that the combination of $\xi \times \kappa^2 = 1.18$
which we have been using in equation (\ref{eq:L_bol}) is reasonable,
to within a accuracy of $\sim \pm 50\%$.

Given the above argument,
let us increase $\xi \times \kappa^2$ by $50\%$, from 1.18 to 1.77.
(If $\xi$ is kept constant, this implies $\kappa=2.55$.)
Because the mass estimate through equation (\ref{eq:L_bol})
is directly proportional to $\xi \times \kappa^2$,
we will then obtain $M_{\rm o} = 14.4^{+1.8}_{-1.3} \ M_\odot$,
and $M_{\rm c} = 6.9^{+2.2}_{-1.9} \ M_\odot$.
Although the mass of the center source becomes consistent with the
mass lower limit in equation (3), that of the off-center source still
remains inconsistent.

Thus, the second problem pointed out in \S~3.3 cannot be solved
despite a series of compromising assumptions described above.
Of course, the problem would be solved if we appleal to more extreme
assumptions, e.g. $D$=3 Mpc.
However, we consider such attempts too artificial.
Furthermore, the issue of too high a disk temperature
(or too low a black-hole mass deirved via $R_{\rm in}$) is found in other ULXs,
including IC~342 source 1 (Okada et al. 1998),
and M81 X-6 (Uno 1997) of which the distance is accurately known
(Freedman et al.\ 1994).
Essentially the same problem has also been reported
from a few Galactic jet sources (Zhang et al. 1997),
of which accurate estimates on $D$, $i$, and the black-hole mass are available.

These arguments suggest that, in some black holes,
the disk temperature can get significantly and systematicaly higher
than is predicted by the standard accretion-disk picture.
Zhang et al. (1997) propose
that such black holes are spinning rapidly,
and the accretion disks are prograde to their rotation;
in such a case, the disk can get closer to the black hole,
and hence get hotter, just as has been observed.
Therefore, the ULXs may be mass-accreting Kerr black holes
with several tens solar masses.
Further examination of this scenario will be presented elsewhere
(Makishima, K., private communication).

\vspace{1pc}\par

Acknowledgement:
We thank Dr.\ T.\ Hanawa and Dr.\ K.\ Nomoto for helpful discussion.

\clearpage

\section*{References}

\re
Burke E. B., Mountain R. W., Daniels P. J., Cooper M. J.,
Dolat V. S.\ 1994, IEEE Trans. Nucl Sci. 41, 375




\re
Dotani T., Inoue H., Mitsuda K., Nagase F., Negoro H., Ueda Y.,
Makishima K., Kubota A. et al.\ 1997, ApJ 485, L87


\re
Fabbiano G.\ 1988, ApJ 325, 544

\re
Fabbiano G.\ 1989, ARA\&A 27, 87


\re
Fleming D. B., Harris W. E., Pritchet C. J., Hanes D. A.\ 1995,
AJ 109, 1044 

\re
Freedman W. L., Hughes S. M., Madore B. F., Mould J. R., Lee M. G.,
Stetson P., Kennicutt R. C., Turner A. et al.\ 1994, ApJ 427, 628

\re
Herreo A., Kudritzki R. P., Gabler R., Vilchez J. M., Gabler A.\ 1995,
A\&A 297, 556

\re
Hummel E., Sancisi R., Ekers D.\ 1984, A\&A 133, 1



\re
Jacoby G. H., Ciardullo R., Harris W. E.\ 1996 ApJ, 462, 1

\re
Kubota A., Tanaka Y., Makishima K., Ueda Y., Dotani T., Inoue H.,
Yamaoka K.\ 1998, PASJ 50, 667

\re
Makishima K., Maejima Y., Mitsuda K., Bradt H. V., Remillard R. A.,
Tuohy I. R., Hoshi R., Nakagawa M.\ 1986, ApJ 308, 635

\re
Makishima K., Ohashi T., Hayashida K., Inoue H., Koyama K., Takano S.,
Tanaka Y., Yoshida A. et al.\ 1989, PASJ 41, 697

\re
Makishima K., Tashiro M., Ebisawa K., Ezawa H., Fukazawa Y., Gunji S.,
Hirayama M., Idesawa E. et al.\ 1996, PASJ 48, 171

\re
Masai K.\ 1984, Ap\&SS 98, 367

\re
Mitsuda K., Inoue H., Koyama K., Makishima K., Matsuoka M., Ogawara Y.,
Shibazaki N., Suzuki K. et al.\ 1984, PASJ 36, 741

\re
Nagase F.\ 1989, PASJ 41, 1

\re
Ohashi T., Ebisawa K., Fukazawa Y., Hiyoshi K., Horii M., Ikebe Y.,
Ikeda H., Inoue H. et al.\ 1996, PASJ 48, 157

\re
Okada K., Dotani T., Makishima K., Mitsuda K., Mihara T.\ 1998, PASJ 50, 25

\re
Petre R., Okada K., Mihara T., Makishima K., Colbert E. J. M.\ 1994,
PASJ 46, L115

\re
Predehl P., Tr$\rm \ddot{u}$mper J.\ 1994, A\&A 290, L29

\re
Read A. M., Ponman T. J., Strickland D. K.\ 1997, MNRAS 286, 626

\re
Reynolds C. S., Loan A. J., Fabian A. C., Makishima K., Brandt W. N.,
Mizuno T.\ 1997, MNRAS 286, 349

\re
Rupen M. P.\ 1991, AJ 102, 1

\re
Shakura N. I., Sunyeav R. A.\ 1973, A\&A 24, 337

\re
Shimura T., Takahara F.\ 1995, ApJ 445, 780

\re
Simard L., Pritchet C. J.\ 1994, AJ 107, 503

\re
Takano M., Mitsuda K., Fukazawa Y., Nagase F.\ 1994, ApJ 436, L47

\re
Tanaka Y., Inoue H., Holt S. S.\ 1994, PASJ 46, L37

\re
Tanaka Y., Lewin W. H. G.\ 1995, in X-ray Binaries, ed W. H. G. Lewin, J. van
Paradijs, W. P. J. van den Heuvel (Cambridge University Press, Cambridge) p126

\re
Tully R. B.\ 1988, Nearby Galaxies Catalogue. Cambridge University Press,
Cambridge 

\re
Ueda Y., Takahashi T., Inoue H., Tsuru T., Sakano M., Ishisaki Y.,
Ogasaka Y., Makishima K. et al.\ 1998, Nature 391, 866

\re
Uno S.\ 1997, PhD Thesis, Gakushuin University

\re
Yamashita A., Dotani T., Bautz M., Crew G., Ezuka H.,
Gendreau K., Kotani T., Mitsuda K. et al.\ 1997, IEEE Trans. Nucl Sci.
44, 847

\re
Volger A., Pietsch W., Kahabka P.\ 1996, A\&A 305, 74

\re
Zang S. N., Ebisawa K., Sunyaev R., Ueda Y., Harmon B. A., Sazonov S.,
Fishman G. J., Inoue H. et al.\ 1997, ApJ 479, 381

\begin{table*}
\begin{center}
Table~1.\hspace{4pt}
Observed count rates (in $\rm 10^{-2}~c~s^{-1}$)
of the two souces in NGC~4565.$^*$ \\
\end{center}
\label{tbl:n4565_projfit}
\vspace{6pt}
\begin{tabular*}{\columnwidth}{@{\hspace{\tabcolsep}
\extracolsep{\fill}}p{6pc}cccc}
\hline 
Source & \multicolumn{3}{c}{ASCA SIS$^\dagger$} & ROSAT HRI$^{\P}$ \\
\hline 
 & soft $^\ddagger$ & medium $^\ddagger$ & hard $^\ddagger$ & \\
\hline\hline 
Off-center   \dotfill & 1.41$\pm$0.10 & 1.24$\pm$0.09 & 0.64$\pm$0.07
& 0.91$\pm$0.13 \\
Center       \dotfill & 0.54$\pm$0.09 & 0.37$\pm$0.07 & 0.27$\pm$0.06
& 0.51$\pm$0.09 \\
ratio$^{\S}$ \dotfill & 2.61$\pm$0.47 & 3.35$\pm$0.68 & 2.37$\pm$0.59
& 1.78$\pm$0.40 \\
\hline
\end{tabular*}
\vspace{6pt}
\par\noindent
$^*$ Errors represent one-sigma stataistical errors.
\par\noindent
$^\dagger$ Count rate of the event inside the rectangle,
obtained through fitting to the one-dimensional profile of
figure 2.
\par\noindent
$^\ddagger$ Soft, medium, and hard energy bands correspond to
0.5--1.5, 1.5--3, and 3--10~keV, respectively.
\par\noindent
$^{\S}$ Count rate ratio of the off-center source to the center source.
\par\noindent
$^{\P}$ Count rate of the events in circular region of radii 12$''$,
derived from ROSAT HRI archival data.
\end{table*}

\begin{table*}
\begin{center}
Table~2.\hspace{4pt}
Results of the joint fits to the SIS and GIS spectra of NGC~4565.$^*$
$^\dagger$ \\
\end{center}
\label{tbl:n4565_specfit}
\vspace{6pt}
\begin{tabular*}{\columnwidth}{@{\hspace{\tabcolsep}
\extracolsep{\fill}}p{10pc}ccc}
\hline\hline 
Model & absorption & $\Gamma$ or $T$ or $T_{\rm in}$ & $\chi^2/\nu$ \\
      &  ($10^{21}$~cm$^{-2}$) & (keV) \\
\hline
Power-law \dotfill                  & 2.4$\pm$0.5 & 1.89$^{+0.09}_{-0.08}$
& 174.2/121 \\
Bremsstraulung \dotfill             & 1.4$\pm$0.4 & 5.6$^{+0.9}_{-0.7}$
& 152.4/121\\
Plasma Emission$^\ddagger$ \dotfill & 1.4$\pm$0.4 & 5.5$^{+0.9}_{-0.7}$
& 152.3/120 \\
Blackbody \dotfill                  & $\le 0.04$     & 0.70$\pm$ 0.02
& 335.4/121 \\
MCD \dotfill                        & $\le 0.2$      & 1.43$^{+0.07}_{-0.06}$
& 145.9/121 \\
\hline
\end{tabular*}
\vspace{6pt}
\par\noindent
$^*$ Spectra of the two sources are co-added together.
\par\noindent
$^\dagger$ Errors represent 90 \% confidence limits.
\par\noindent
$^\ddagger$ Using the Masai (1984) code. The abundance has been constrained
to be $\le 0.1$ solar.
\end{table*}

\begin{table*}
\begin{center}
\vspace{6pt}
Table~3.\hspace{4pt}
Estimates of the spectra of individual sources.\\
\end{center}
\caption{Estimates of the spectra of individual sources.}
\label{tbl:n4565_specfit2}
\vspace{6pt}
\begin{tabular*}{\columnwidth}{@{\hspace{\tabcolsep}
\extracolsep{\fill}}p{6pc}ccc}
\hline\hline 
Source & absorption            & $T_{\rm in}$  & bolometric flux  \\
       & (10$^{21}$ cm$^{-2}$) & (keV)         & ($^\dagger$) \\
\hline 
Off-center \dotfill & $\le$0.2 & 1.39$\pm0.08$ & 1.82 \\
Center     \dotfill & $\le$0.5 & 1.59$^{+0.32}_{-0.23}$ & 0.72 \\
\hline 
\end{tabular*}
\vspace{6pt}
\par\noindent
$^*$ The SIS/GIS spectra of NGC~4565 were fitted jointly with two MCD
components,
which are constrained to produce the center and off-center source spectrum.
\par\noindent
$^\dagger$ In unit of $10^{-12}$ ergs s$^{-1}$ cm$^{-2}$.
\end{table*}

\clearpage
\centerline{Figure Captions}
\bigskip

\begin{fv}{1}{}
{X-ray images of NGC~4565, superposed on the optical
(Digital Sky Survey) image (J2000 coordinates).
(a) The ROSAT HRI image.
(b) The ASCA SIS image in the 0.5--10 keV band.
It was smoothed with a Gaussian distribution
of $\sigma =0.\hspace{-2pt}'1$.}
\end{fv}

\begin{fv}{2}{}
{
The projected one-dimensional X-ray brightness distribution of
NGC~4565, in 0.5--1.5 keV (panel~a), 1.5--3 keV (panel~b),
3--10 keV (panel~c), and 0.5--10 keV (panel~d).
The SIS events within the rectangle of
figure 1b are used.
The dotted, dashed, and dot-dashed lines indicate
the model for the center-source,
that for the off-center source, and a constant background, respectively.
The solid histgrams show the sum of threse three model components,
to be compared with the data.
}
\end{fv}

\begin{fv}{3}{}
{The SIS and GIS spectra of NGC~4565, dominated by the two sources.
The histgram shows the best fit model and the crosses represent the
observed spectra.
(a)~A fit with a single MCD model.
(b)~A fit with two MCD models, incorporating the constraints
that each model can simultaneously reproduce the three-band spectrum
(data points with wide bin) of the corresponding source.}
\end{fv}

\begin{figure}[h]
\centerline{\hbox{
\psfig{file=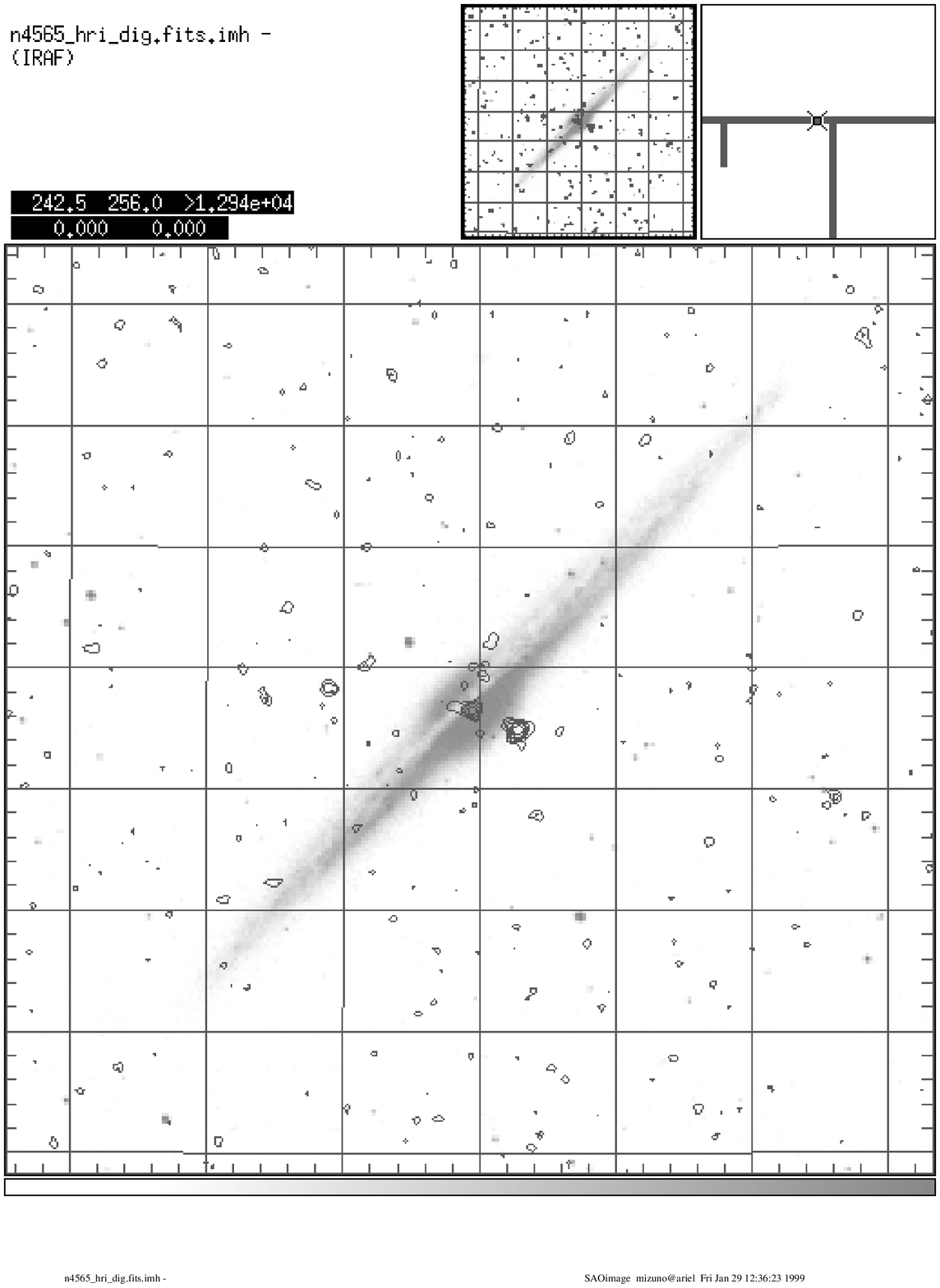,height=7.1cm,width=8cm,rheight=0cm,clip=}
\psfig{file=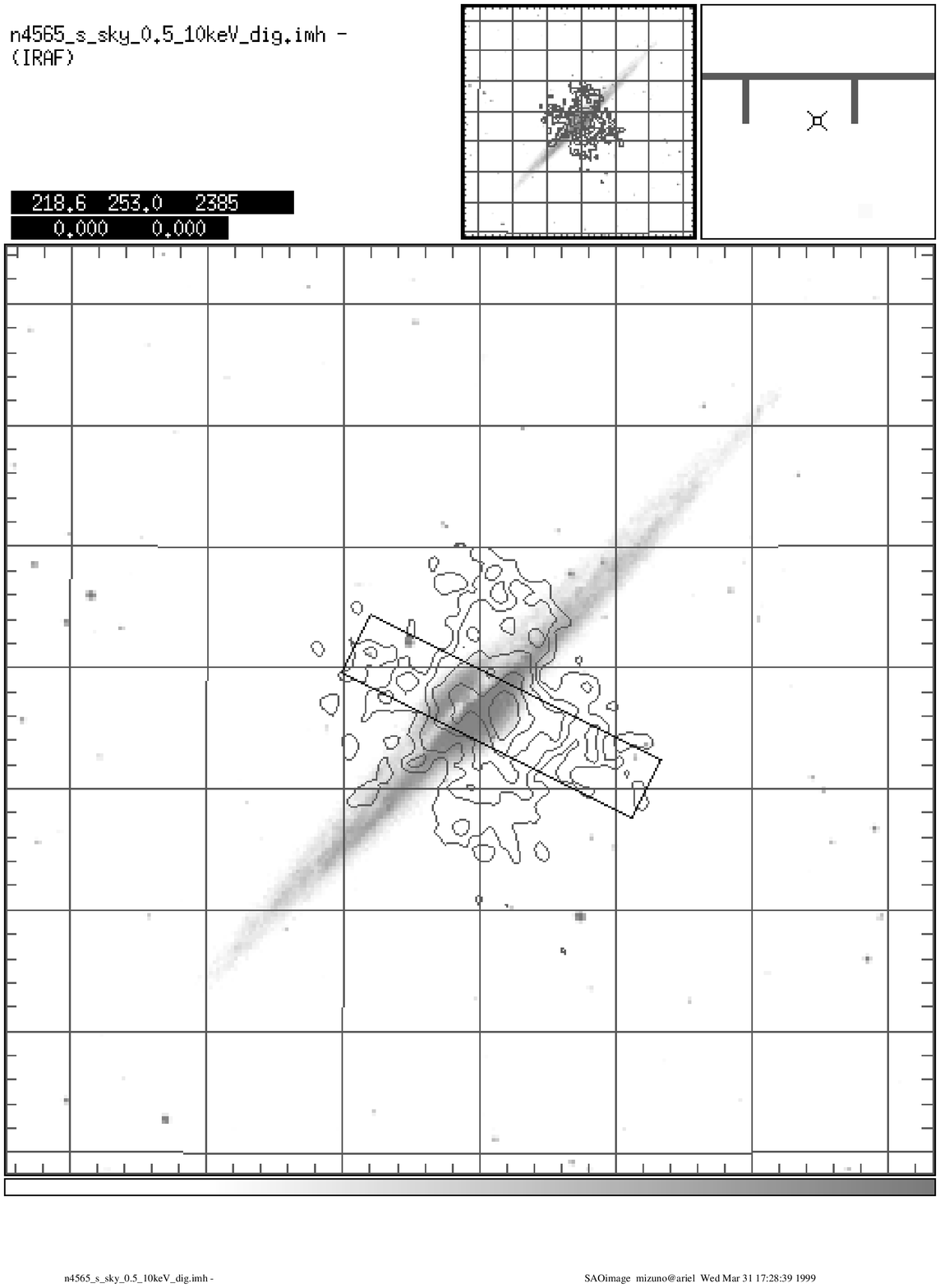,height=7.1cm,width=8cm,rheight=0cm,clip=}
}}
\setlength{\unitlength}{1mm}
\begin{picture}(80,71)(0,0)
\put(-0.5,0.5){\thicklines\framebox(79.5,70.5)}
\put(81,0.5){\thicklines\framebox(79.5,70.5)}
\put(64,-5){$12^{\rm h}36^{\rm m}00^{\rm s}$}
\put(39,-5){$20^{\rm s}$}
\put(6,-5){$40^{\rm s}$}
\put(-12,64){$04^{'}$}
\put(-12,36){$26^{\circ}00^{'}$}
\put(-12,8){$56^{'}$}
\put(71,60){\large(a)}
\put(152,60){\large(b)}
\end{picture}
\label{fig:n4565_image}
\end{figure}

\vspace{4cm}

\centerline{Figure 1}

\newpage

\begin{figure}[htbp]
\centering
\centerline{\hbox{
\psfig{file=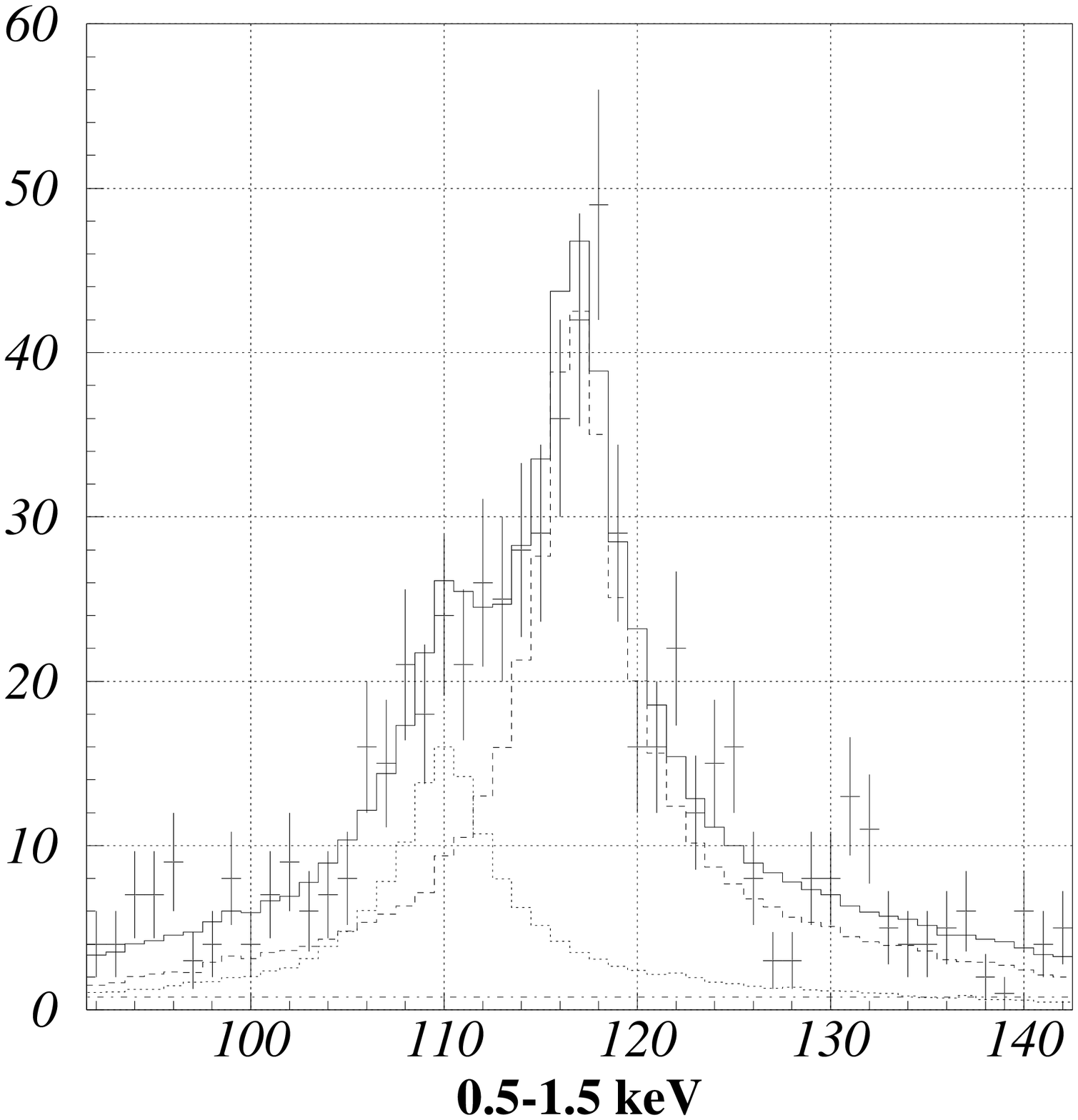,height=6cm}
\psfig{file=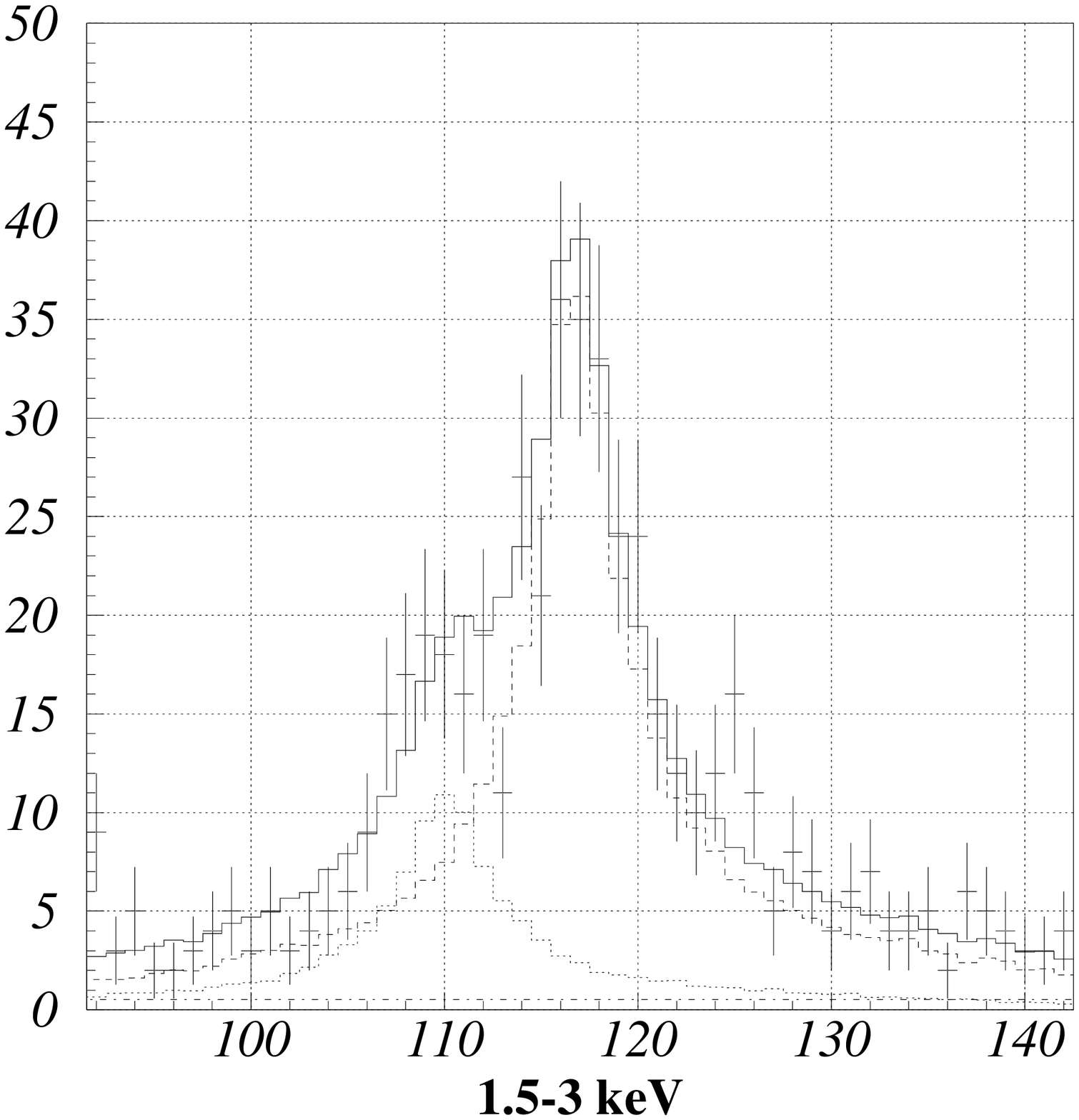,height=6cm}
}}
\centerline{\hbox{
\psfig{file=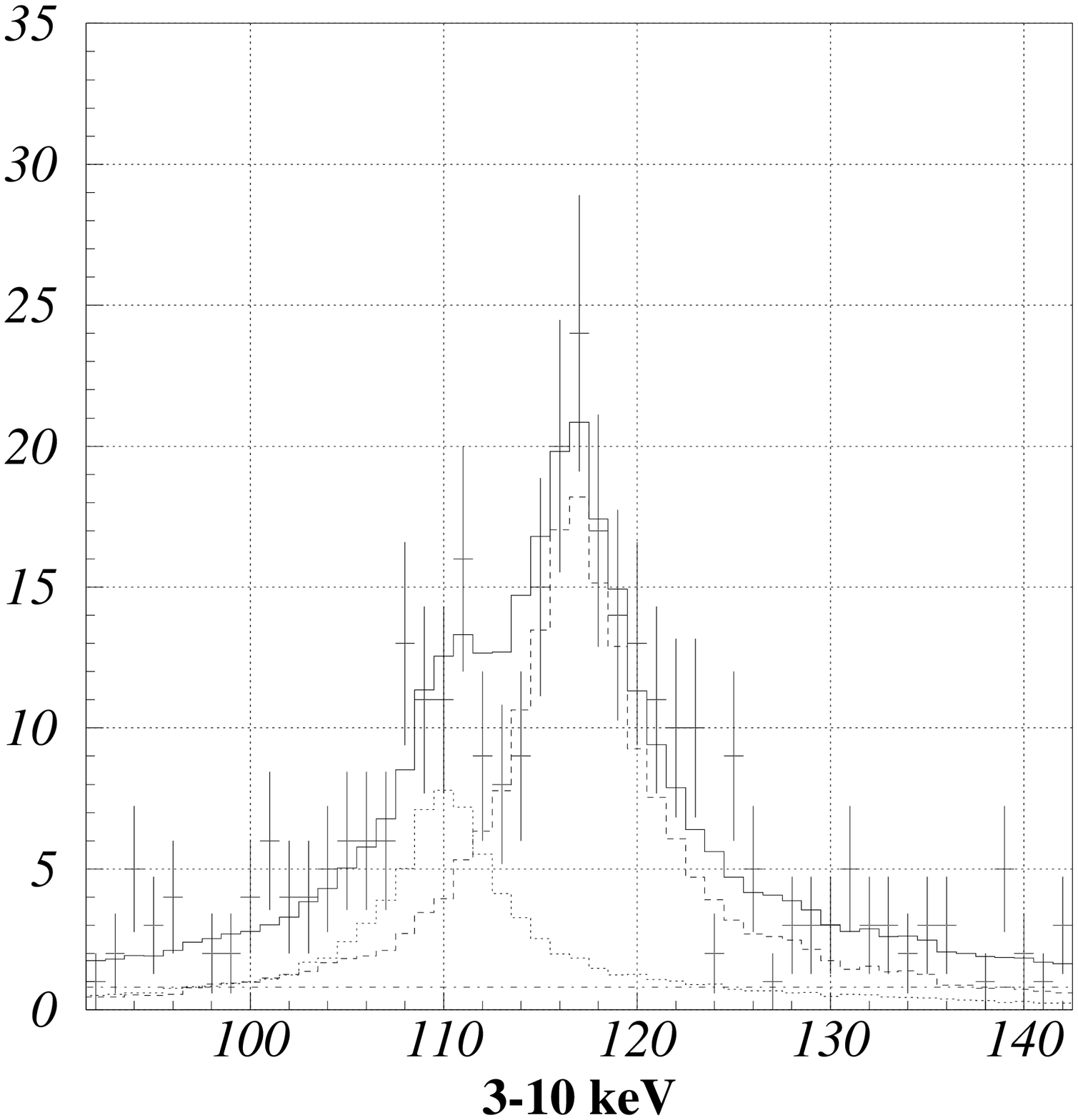,height=6cm,rheight=0cm,clip=}
\psfig{file=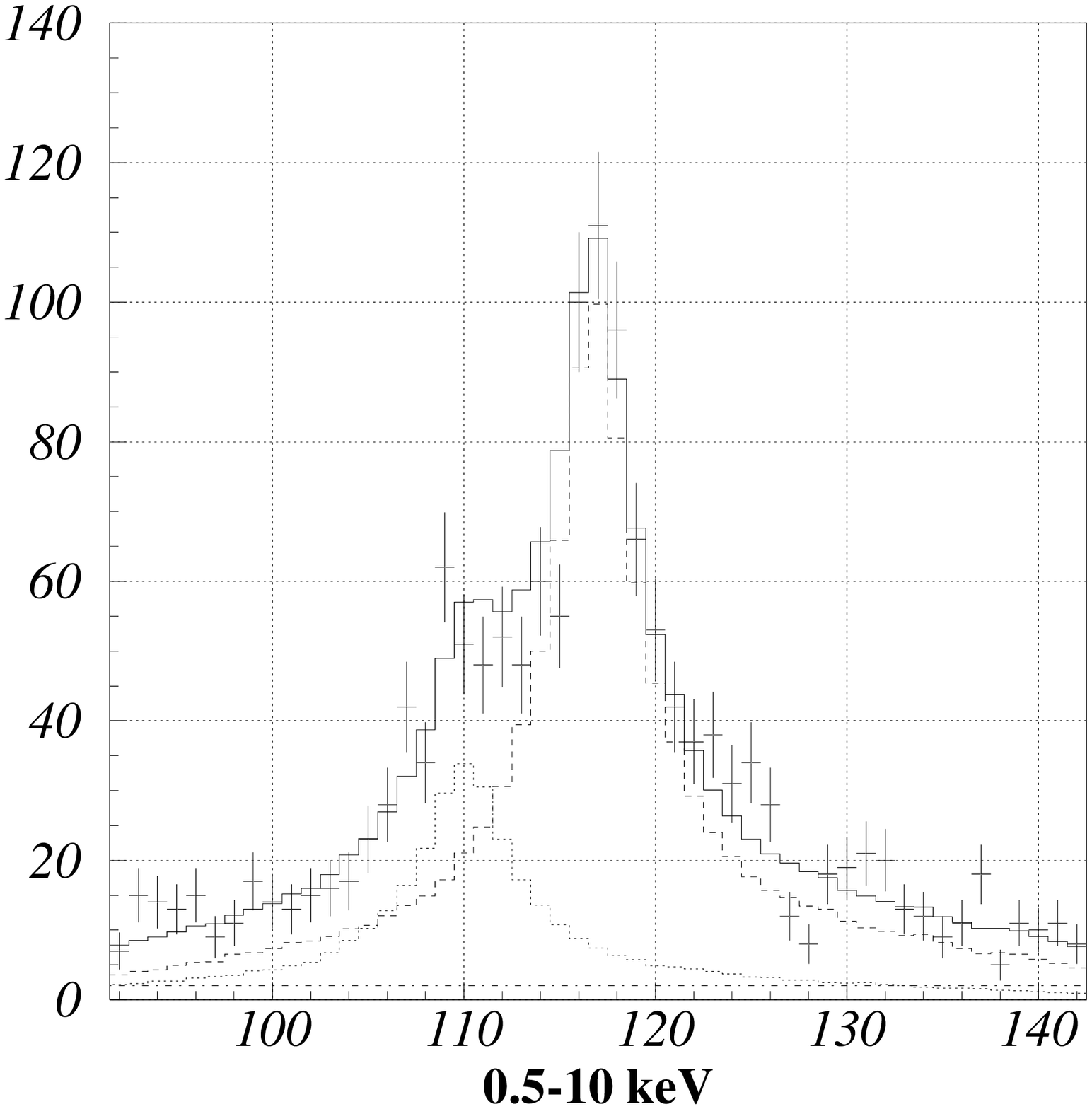,height=6cm,rheight=0cm,clip=}
}}
\setlength{\unitlength}{1mm}
\begin{picture}(50,50)(0,0)
\put(10,95){\large(a)}
\put(70,95){\large(b)}
\put(10,30){\large(c)}
\put(70,30){\large(d)}
\put(13.7,88){\vector(1,0){5.3}}
\put(13.7,88){\vector(-1,0){4.2}}
\put(12.4,83){\large $1'$}
\end{picture}
\label{fig:n4565_projfit}
\end{figure}

\vspace{4cm}

\centerline{Figure 2}

\newpage

\begin{figure}[h]
\centerline{\hbox{
\psfig{file=diskbb_fit.ps2,angle=-90,height=7cm,rheight=0cm,clip=}
\psfig{file=2source_diskbb_fit.ps2,angle=-90,height=7cm,rheight=0cm,clip=}
}}
\setlength{\unitlength}{1mm}
\begin{picture}(70,70)(0,0)
\put(60,50){\large(a)}
\put(155,50){\large(b)}
\end{picture}
\label{fig:n4565_specfit}
\end{figure}

\vspace{4cm}

\centerline{Figure 3}

\end{document}